\documentclass[11pt]{article}
\usepackage{graphicx}
\usepackage{color, rotating, amsmath, amsfonts , subfig , color, graphicx}
\usepackage[justification=centering]{caption}

\usepackage{lineno,hyperref}
\usepackage{setspace}

\usepackage[mathscr]{euscript}

\newcommand{\vsp}{\vspace*{3mm}}

\newcommand{\be}{\begin{equation}}
\newcommand{\ee}{\end{equation}}
\newcommand{\bd}{\begin{displaymath}}
\newcommand{\ed}{\end{displaymath}}
\newcommand{\bea}{\begin{eqnarray}}
\newcommand{\eea}{\end{eqnarray}}
\newcommand{\bean}{\begin{eqnarray*}}
\newcommand{\eean}{\end{eqnarray*}}
\begin{document}

\begin{center}{\LARGE\sc  Improved resection margins in breast-conserving surgery using Terahertz Pulsed imaging data}\end{center}
\begin{center}{\sc A Santaolalla \footnote[2]{Translational Oncology and Urology Research (TOUR), School of Cancer and Pharmaceutical Sciences, King's College London, London SE1 9RT, UK. Email: aida.santaolalla@kcl.ac.uk}\, \footnote[5]{Equally contributing primary authors}, M Sheikh\footnote[3]{Institute for Mathematical and Molecular Biomedicine, Kings College London, London SE1 1UL, UK. Email: mansoor.sheikh@kcl.ac.uk}\,\footnotemark[5], 
M Van Hemelrijck\footnotemark[2], A Portieri\footnote[4]{TeraView Ltd., Cambridge, CB4 0WS, UK}, ACC Coolen\footnotemark[3] \\
\today}\end{center}
\vsp

\begin{abstract}
New statistical methods were employed to improve the ability to distinguish benign from malignant breast tissue ex vivo in a recent study. The ultimate aim was to improve the intraoperative assessment of positive tumour margins in breast-conserving surgery (BCS), potentially reducing patient re-operation rates. A multivariate Bayesian classifier was applied to the waveform samples produced by a Terahertz Pulsed Imaging (TPI) handheld probe system in order to discriminate tumour from benign breast tissue, obtaining a sensitivity of 96\% and specificity of 95\%. 

We compare these results to traditional and to state-of-the-art methods for determining resection margins.  Given the general nature of the classifier, it is expected that this method can be applied to other tumour types where resection margins are also critical.
\end{abstract}





\section{Introduction}

Breast cancer is the most frequent cancer among women worldwide \cite{fitzmaurice2017global} and the leading cause of cancer death in females, accounting for 23\% of the total cancer cases and 14\% of the cancer deaths \cite{jemal2011global}. Surgery to remove the primary tumour is one of the main curative treatments in breast cancer which has evolved over time from radical surgery to more conservative approaches, such as  breast conserving surgery \cite{franceschini2008conservative} \cite{sakorafas2001breast}.

\begin{figure}[h]
\centering
	\includegraphics[height=5cm, trim={0 5cm 0 5cm}]{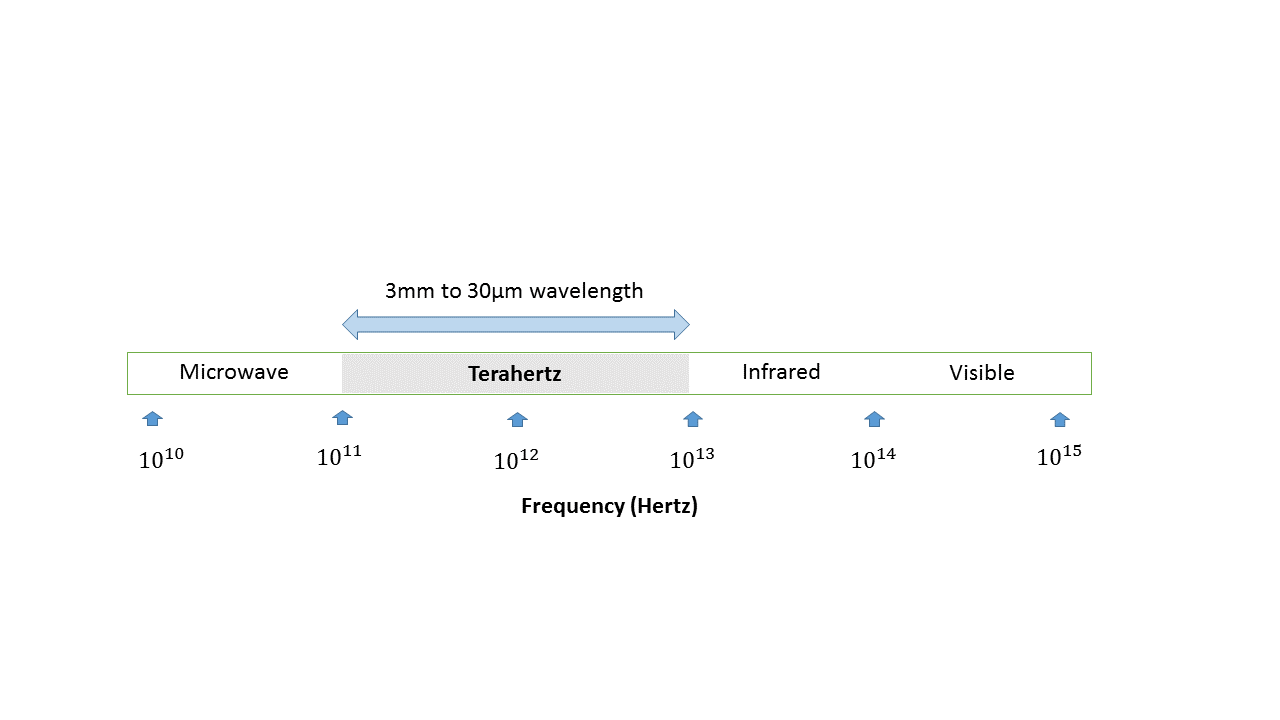} 
	\caption{\small Electromagnetic spectrum showing frequency of Terahertz radiation}
	\label{fig:spectrum}
\end{figure}

Good surgical outcomes are related to the accuracy in the resection of the tumour margins during surgery. Currently, a significant number of patients need to undergo additional surgery to obtain clear margins and/or remove remaining lymph nodes in the axilla, causing a significant physical and psychological morbidity in patients  \cite{fitzal2006breast}. To date, histopathology, which occurs after surgery, is the gold standard to determine tumour margins. Hence, new techniques are being investigated to assess tumour margins intraoperatively. Terahertz (THz) radiation is one such tool (see figure \ref{fig:spectrum}); it does not produce harmful ionization in biological tissues and therefore has been applied in cancer studies previously \cite{fitzgerald2006terahertz}.

In order to implement this technology to assess tumour margins and sentinel lymph nodes in breast cancer patients intraoperatively, Teraview Ltd.(Cambridge, UK) developed a handheld THz Probe to be used in breast conserving surgery, eventually aiming to reduce the number of reoperations.

In \cite{fitzgerald2006terahertz}, a feasibility study was performed to determine the capability of the  handheld THz Probe to distinguish between different breast cancer tissues ex vivo.The THz raw data obtained was complex, multidimensional and hold lots of noise. There is a lack of standardised procedures for data acquisition, post processing, image analysis and interpretation of the data generated with this technology. Therefore, \cite{fitzgerald2006terahertz} used two different statistical approaches to discriminate between different breast tissue samples based on Terahertz pulsed imaging (TPI) data produced by the Teraview probe. 

In the current paper, we aim to improve the prediction capabilities of the classifier by utilising a novel Bayesian classifier \cite{sheikh2017accurate} which improves tissue discrimination by distinguishing between tumour, fibrous and adipose tissue in comparison with other technologies employ to assess tumour margins and in comparison with the previous feasibility study.

The large number of covariates in the dataset along with a relatively small number of tissue samples makes the prediction of the clinical outcome difficult because of overfitting problems and prohibitive computation demands. Typically Bayesian calculations rely on computational methods such as Markov Chain Monte Carlo to compute integrals. In our approach, this is overcome by judicious model selection resulting in the integrals which scale with dimension to be analytically tractable. 

The paper is organised as follows. Section \ref{sec:data} describes the data collection and pre-processing methods. Section \ref{sec:classify} outlines the new Bayesian classifier used for determining tissue class. The results are stated in section  \ref{sec:results}. We review results for existing and state-of-the-art technologies for this clinical problem in section \ref{sec:compare}. Discussion and next steps are in section \ref{sec:conclude}.  

\section{Data collection}
\label{sec:data}

\subsection{Technology}
\begin{figure}[h]
\centering
	\includegraphics[height=5cm]{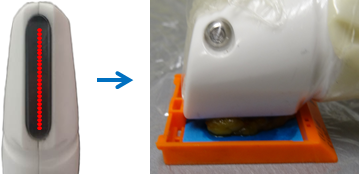} 
	\caption{\small Schematic description of the raw data acquisition. Images are courtesy of the study (REC 12-EE-0493). TPI handheld probe measurement of tissue sample positioned in histology cassette. Residual THz pulses are received by each pixel from the tissue producing typical TPI waveforms per pixel.}
	\label{fig:probe}
\end{figure}

The handheld terahertz pulsed imaging (TPI) probe device (Teraview)  has 26 pixels \ref{fig:probe}. Each pixel transmits terahertz electromagnetic radiation of frequency 0.1-2.0 THz  to the biological sample and records any residual signal transmitted from the tissue. Technical details can be found in \cite{grootendorst2017use}.

\subsection{Data acquisition}
Imaging data was produced from scanning 46 freshly excised breast cancer samples obtained from 30 breast cancer patients treated at Guy's and St Thomas' NHS Foundation Trust (GSTT) following a breast conservation surgery or mastectomy between August 2013 and August 2014. This dataset was acquired as part of a feasibility study to test the utility of the handheld (TPI) probe to discriminate between breast tissue types (REC 12-EE-0493).

The ex vivo breast tissue samples contained different proportions of adipose, fibrous and tumour tissue. Each sample was assessed for detailed histopathology by two experts blinded for the patient details that labelled the percentage of adipose, fibrous and tumour at  pixel  level for each of the samples. Following the histopathology assessment, the samples were scanned with the probe, using air as a reference, to produce the data for the study. 
Each pixel transmitted many pulses to the tissue sample and received the residual data. These were averaged within the device to produce a waveform which was the input for the statistical analysis. Ideally if the complete amount of the residual information from each pulse was preserved, a more accurate idea of the signal variability could have been obtained.

\begin{figure}[h]
\centering
	\includegraphics[height=9cm]{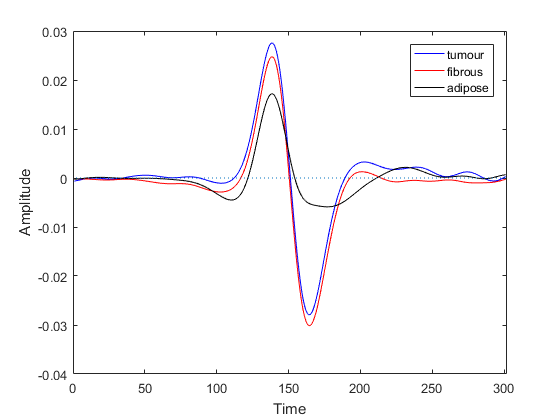} 
	\caption{\small For each tissue type present in the sample,  the TPI waveform presented a different shape: tumour (blue), fibrous (red) and adipose cells (black). }
	\label{fig:waveform}
\end{figure}

An averaged waveform with 674 data points was produced by each pixel. It was determined that the outermost points were either zero or noise so these were removed \cite{grootendorst2017use}. This reduced the number of data points to 301 per waveform (see figure \ref{fig:waveform}). The detailed histopathology results were appended to the Terahertz (THz) data for each sample. In the language of statistical inference, this is a supervised learning problem with the predictor being the waveform and the response variable being the histopathology result. The size and orientation of the samples were varied. This resulted in varying numbers of pixels per tissue. The final dataset included a total of 257 pixels resulting in a $257 \times 301$ matrix of real-valued data.

\subsection{Data pre-processing}
Our aim is to discriminate each impulse function (waveform) per pixel per sample into tumour, fibrous or benign breast tissue. However it is difficult to do so visually (see figure \ref{fig:waveform}). Statistical analysis is therefore required.

\noindent \emph{Feature selection}. Given the high dimensionality of the data, some methods use feature selection to choose the most informative features. In \cite{grootendorst2017use}, the Support Vector Machine methodology used heuristic methods to determine which features had the greatest effect on classification results. For the Naive Bayesian classifier, also used in \cite{grootendorst2017use}, overfitting is not as large a concern. Gaussian deconvolution, common in image processing, was used before sending the data to the classifier. In this present paper, we found that classification based on the raw waveform data produced the best results.

Throughout this paper, we distinguish between a tissue sample which is a physical sample from a patient and a data sample understood in the statisical sense. In this study there were 46 tissues samples collected which produced 257 data samples.

\section{Classification methods}
\label{sec:classify}
\subsection{General method}
A multivariate Bayesian classification algorithm \cite{sheikh2017accurate} was applied to the TPI waveform data to discriminate benign from malignant breast tissue. The classifier used supervised learning techniques with the class labels derived from the histopathologically defined tissue content. The trained classifier was used to predict the classes of new observations. 

In statistical learning, a classifier is a function which maps a data point to a class prediction. There are four stages to a classification problem. The first is model selection. We use the framework outlined in \cite{sheikh2017accurate}. The second stage is to train our chosen model on the training data. In our case, this involves learning a set of parameters for each class: tumour, fibrous and adipose.  In the third stage, the classifier uses a probabilistic approach assuming that the observations in each class were generated by a multivariate Gaussian distribution specific to that class. Predictive probabilities of a new data sample belonging to a certain class are calculated. Lastly, decision theory is applied to translate those probabilities into class predictions. This is done by simply choosing the class with the highest probability conditioned on the data sample. While traditional statistical methods produce probability point estimates, probability distributions are produced in our Bayesian approach. This allows us to quantify the uncertainty in the predictions made.

The performance of the classifier was evaluated using leave-one-out cross-validation (LOOCV). Training of the Bayesian classifier occurred by leaving out a single data sample and learning parameters from the remaining data samples. The estimated parameters were then used to classify the data sample that was left out. This process was repeated for all the samples. The results were compiled to estimate accuracy, sensitivity, specificity, positive predictive value (PPV) and negative predictive value (NPV) to distinguish malignant from benign tissue.   

\vsp

\noindent \emph{Overfitting}. Typically, data with more covariates, $d$, than samples, $n$, leads to overfitting \cite{coolen2017replica}. This is the case with our data where $n=257$ and $d=301$. This is somewhat mitigated by using a fully Bayesian classification approach \cite{sheikh2017accurate}.  Inference methods such as maximum likelihood work well when $n \gg d$ but are prone to overfitting when $n \leq d$. The classifier used in this paper avoids taking point estimates of unknown parameters. Instead it treats them in a Bayesian way by assuming they are random variables and integrating over them. By carefully choosing the prior probability distributions, a closed-form for the predictive probabilities is derived. The final estimation of hyperparameters is done analytically. An added advantage is that computationally expensive cross-validation methods are avoided allowing us to handle larger data-sets.

\subsection{Application to Terahertz Pulsed Imaging}
\label{sec:scenarios}
The data could be categorised under three different scenarios depending on the particular research question assessed. The class label for each sample will be determined by:

\begin{itemize}
\item Scenario 1:  pixels were marked as tumour when containing any amount of cancer cells (if the tumour percentage was greater than zero, the TPI impulse function was considered as tumour, irrespective of the content of fibrous or adipose tissue), otherwise the tissue content of the pixel was defined by the highest percentage of fibrous or adipose tissue; 
\item Scenario 2: pixels were marked as tumour when containing any amount of cancer cells (if the tumour percentage was greater than zero, the TPI function was considered as tumour, independently of the content of fibrous or adipose tissue), otherwise tissue was considered to be “benign”. In this scenario, adipose and fibrous tissues were grouped together; 
\item Scenario 3: pixels were marked based on their tissue content, so the tissue content of each pixel was defined by the highest percentage. 
\end{itemize}

Given that the main aim of the project was to discriminate malignant tissue from benign breast tissue, and that benign breast tissue could be subdivided into fibrous and adipose tissue respectively, scenario 1 was the preferred option in the study. This results in class sizes of 115, 100 and 42 (totalling 257 data samples).

The original THz time domain pulses were assessed in their ability to discriminate tumour from healthy (fibrous/adipose) breast tissue using our probabilistic Bayesian classification algorithm \cite{sheikh2017accurate}. 

\vsp

\noindent \emph{Class imbalance problem}. Classification methods with imbalanced class sizes tend to favour the majority class \cite{wallace2011class}. This is a well known phenomenon in classification literature. The problem frequently occurs in medical data-sets where the rare or diseased cases have many fewer samples. In our case, the data collected focussed on tumorous tissue and hence the adipose class has many fewer samples. 

Methods to compensate for this effect can be grouped into pre-processing methods or algorithmic changes. Pre-processing can be achieved by over-sampling the minority class \cite{chawla2002smote} or under-sampling the majority class \cite{drummond2003c4}. On the other hand, the classification algorithm itself can be changed \cite{wallace2011class}. An example is to specify a cost function for misclassification to mitigate the class imbalance effect. We have not attempted these methods in this paper.

\section{Results}
\label{sec:results}
Classification accuracy is first displayed for the three class analysis (tumour, fibrous and adipose) is in Table \ref{tab:confusion}.

\begin{table}[h!]
\centering 
\begin{tabular}{c c c c} 
\hline\hline 
  &  Predict 1 & Predict 2 & Predict 3   \\ [0.5ex] 
\hline 
True class 1 & 110 & 5 & 0 \\ 
True class 2 & 5 & 95 & 0 \\ 
True class 3 &  2& 27 & 13  \\ 
\hline \hline 
\end{tabular}
\caption{\small Confusion matrix. The classification results are split into three classes: class 1 represents tumour, class 2 fibrous and class 3 adipose tissue. Conventions for the confusion matrix are described in appendix \ref{app:confusion}} 
\label{tab:confusion}
\end{table}

The number of corrected predicted samples was 218 out of 257 which gave a classification accuracy of 85\%. However the clinically important decision was whether the sample contains tumour or not.  Classes 2 and 3 represented benign tissue. By aggregating these benign classes, we obtained the confusion matrix in Table \ref{tab:aggregate}. 

For the two class results, the classification accuracy, sensitivity and specificity were 95\%, 96\% and 95\%.  In addition the PPV and NPV values were 94\% and 96\%.

\begin{table}[h]
\centering 
\begin{tabular}{c c c} 
\hline\hline 
  &  Predict 1 & Predict 2  \\ [0.5ex] 
\hline 
True class 1 & 110 & 5  \\ 
True class 2 & 7 & 135  \\ 
\hline \hline 
\end{tabular}
\caption{\small Aggregated confusion matrix. Classes 2 and 3 representing fibrous and adipose tissue have been aggregated into one class in line with the clinical objective.} 
\label{tab:aggregate}
\end{table}

Next we summarise the classification results on the same data-set using our method and those from previous studies \cite{grootendorst2017use} (see table \ref{tab:compare}).

\begin{table}[h]
\centering 
\begin{tabular}{c c c c} 
\hline\hline 
Classifier  &  Bayesian method  & Support Vector Machine \cite{grootendorst2017use}&  Naive Bayes \cite{grootendorst2017use}\\ [0.5ex] 
\hline 
Accuracy  & {\bf 95\%} &  75\%  & 69\% \\ 
Sensitivity & {\bf 96\%} &  86\%  &  89 \%\\ 
Specificity & {\bf 95\%}  &  66\%  &  53 \%  \\ 
PPV & {\bf 94\%} &  67\%   &    60  \%  \\ 
NPV & {\bf 96\%}  &  85\%   &   86 \%  \\ 
\hline \hline 
\end{tabular}
\caption{\small Comparison of classification results. The \emph{Bayesian method} column is the result of this paper, \emph{Support Vector Machine} and \emph{Naive Bayes} are the support vector machine classifier and Naive Bayes methods of \cite{grootendorst2017use}. Figures in bold are the best for that metric. } 
\label{tab:compare}
\end{table}

\section{Comparison to existing methods }
\label{sec:compare}
Different intraoperative margin assessment (IMA) tools are being investigated to assess tumour resection margins. Some of these techniques are clinically established, such as frozen section analysis \cite{cendan2005accuracy}\cite{kim2016efficacy}\cite{jorns2012intraoperative}, specimen radiography, intraoperative ultrasound, touch imprint cytology and optical spectroscopy. However these IMAs present diverse performance and limitations in terms of accuracy, speed, cost, and reliability. Therefore, new IMA tools are currently under development including Raman spectroscopy \cite{haka2005diagnosing}\cite{liu2013resonance}, microcomputed CT, mass spectroscopy  \cite{st2017rapid}, radiofrequency spectroscopy\cite{schnabel2014randomized} or fluorescence imaging. 

Our classification results were materially better than the performance of IMAs currently used, as reported by a recent meta-analysis \cite{st2017diagnostic} (see table \ref{tab:other_method}).

\begin{table}[h]
\centering 
\begin{tabular}{c c c c c} 
\hline\hline 
  &  Number of patients/samples & Accuracy & Sensitivity & Specificity  \\ [0.5ex] 
\hline 
Frozen section analysis (FSA)  & 46-1327 & 84-98 \% & 78-91\% & 92-98\%\\ 
Specimen radiography & 12-119  & 33-84\% & 45-61\% & 77-89\% \\ 
Intraoperative ultrasound  & 81-225  & 62-80\% & 36-79\% & 66-91\% \\ 
Touch imprint cytology  & 27-510  & 78-99\% & 71-97\% & 90-98\% \\ 
Optical spectroscopy  & 20-179  & 75-94\% & 74-91\% & 65-96\% \\ 
Support Vector Machine & 257 & 75\% & 86\% & 66\% \\
Naive Bayesian method & 257 & 69\% & 89\% & 53\% \\
New Bayesian method  & 257 & 95\% & 96\% & 95\% \\
\hline \hline 
\end{tabular}
\caption{Comparison to existing methods. These results are taken from \cite{st2017diagnostic}. The number of patients/samples and accuracy are the ranges across all studies reviewed. The sensitivity and specificity are the result of their meta-analysis.} 
\label{tab:other_method}
\end{table}

\section{Discussion and next steps}
\label{sec:conclude}
Our results are comparable to Frozen Section Analysis, which is considered the gold standard. Therefore our classifier is learning most of the signal in the data.  Comparing to previous methods, our classification method performs better on all measures than the previous classifiers from \cite{fitzgerald2006terahertz} on the given data-set. The Support Vector Machine method used heuristics to select suitable features. Our method has the advantage of using all the available data and requiring no heuristic choices. Additionally many classification methods require ad-hoc choices for the tuning parameters (hyperparameters). In our case, this hyperparameter optimisation is analytic.

The data-set used has class sizes of 115, 100 and 42. This class imbalance will lead to worse classification than a balanced data-set. We plan to use data pre-processing methods to help with class imbalance. Given this involves multiple runs with different samples removed, it requires much more time than no pre-processing.

\subsection{Next steps}

Three tumour types are present in the data (invasive ductal carcinoma, invasive lobular carcinoma and invasive tubular carcinoma). It is likely that these will have statistically different TPI waveforms.  Our method labels all three tumour types as one class despite being able to handle many more than three classes. However too few data samples relative to the number of covariates leads to insufficient characterisation of each class. If sufficient numbers of each type were present in the data, it is likely that classification accuracy would improve. 

The composition of human breast tissue can vary with patient age. Having access to additional data such as patient age is likely to improve classification accuracy.

Our method of leave-one-out cross-validation is a standard statistical learning procedure given small datasets. It was used to allow comparison to existing literature, specifically \cite{grootendorst2017use}. However it is recognised that including data samples from the same patient is not clinically realistic. This likely flattered our classification accuracy metrics. This should be compensated by increasing the overall training size.

Our data was obtained from ex vivo tissue samples. Since Terahertz radiation is non-ionising, it would be interesting to re-run the analysis using data from in vivo clinical trials.

Lastly, our analytical methods are sufficiently general to be applied to other tumour types.

\section{Acknowledgements}
We are grateful for the use of data produced by the handheld THz Probe developed by Teraview Ltd (Cambridge, UK) in a collaborative project with Professor Purushotham and his PhD student Maarten Grootendorst. The work of Mansoor Sheikh was supported by BBSRC and GSK Ltd.

\appendix

\section{Confusion matrix terminology}
\begin{table}[h]

\centering 
\begin{tabular}{c c c} 
\hline\hline 
  &  Predict 1 & Predict 2  \\ [0.5ex] 
\hline 
True class 1 & a & b  \\ 
True class 2 & c & d  \\ 
\hline \hline 
\end{tabular}
\caption{Aggregated confusion matrix} 
\label{app:confusion}
\end{table}

\noindent $a=$True Positive (TP),    $b=$False Negative (FN),    $c=$False Positive (FP),     $d=$True Negative (TN)
\vsp

\noindent Accuracy $=\frac{a+d}{a+b+c+d}$, Sensitivity $=\frac{a}{a+b}$, Specificity $=\frac{d}{c+d}$, PPV$=\frac{a}{a+c}$ and NPV$=\frac{d}{b+d}$.

\appendix

\end{document}